\documentclass{article} 
\begin{document} 
\begin{center} 
\title{Observers and the AdS/CFT correspondence} 
\author{M.Dance}
\maketitle 
\end{center}  
 
\begin{abstract} 
The paper~\cite{Dance0601} tentatively suggested a physical picture that might underlie string theories. The string parameters $\tau $ and $\sigma_i $ were interpreted as spacetime dimensions which a simple quantum mechanical observer can observe, while symmetries of the relevant observer states could limit the observability of other dimensions.  The present paper extends the discussion by outlining how consideration of different observers, at least one of which is quantum mechanical, might provide insights into the nature of the AdS/CFT correspondence. It is suggested that such correspondences might arise as different forms of theories about the world as perceived by different observers.
\end{abstract} 

\section{Introduction}

Heisenberg noted that his uncertainty principle applies to observers as well as to observed systems~\cite{heisenberg}. He postulated that any practical effect of observer indeterminacy could be eliminated by allowing the observer's mass to approach infinity, but gravity would then become important.  To arrive at a theory that unifies quantum mechanics and gravity, it might therefore be useful to include quantum mechanical observers.  

Physically, non-commutative geometries might incorporate quantum mechanical uncertainty in the position and momentum of an observer's centre of mass.  For example, Bander~\cite{Bander} has shown that non-commutative geometry is an effective low-energy theory of systems coupled to an auxiliary system; one might interpret the auxiliary system of~\cite{Bander} as an observer.  
 
It then seems possible that a richer theory remains to be uncovered, one which includes properties of a quantum mechanical observer's states. In~\cite{Dance0601}, it was suggested that a quantum mechanical observer term in a Lagrangian density corresponds to a string theory if symmetries of relevant observer states make corresponding spacetime dimensions unobservable to the observer, and if the radial dimension $r$ is not observable by such an observer, requiring instead an intelligent observer. These ideas were discussed in particular for atomic quantum mechanical observers.  \cite{Dance0601} suggested that this physical picture might even underlie string theories. An atomic observer was the focus of the discussion. The paper~\cite{Dance1011} extended the discussion to a molecular observer, and noted that the symmetries of molecular bonds might effectively give rise to string theories (and theories including branes) in the same way as in~\cite{Dance0601}.  These arguments do not in fact require an observer term in the Lagrangian density; the term could instead describe a field in the external world. The key element is the inclusion of a transformation between the coordinates of different observers' reference systems, and the treatment of this transformation term as being of dynamical interest in its own right.
 
Recent articles of interest in the area of quantum reference frames include e.g.~\cite{Bartlett0610}, \cite{Wang0609}, \cite{Wang0806}, \cite{Dance0610}, \cite{Dance0905} and~\cite{Dance1011}. 

The present paper summarises some previous work including~\cite{Dance0601} and outlines how consideration of quantum mechanical observers might provide insights into the nature of the AdS/CFT correspondence.

\section{Summary of the physical picture}

This section briefly summarises previous work in e.g.~\cite{Dance0601}.  To incorporate observer dynamics, a simple observer term was added in~\cite{Dance0601} to a field theory Lagrangian density in a flat classical background spacetime. The term represented the dynamics of the centre of mass of the quantum mechanical observer O1 (but could in fact stand for a representative point in an external observed system):-

\begin{equation}
L^{EK}_{obs} = \frac{1}{2}m \eta_{ij}\frac{dX^i}{dT}\frac{dX^j}{dT}
\end{equation}
which corresponds to a Lagrangian density
\begin{equation}
\mathcal{L} =
\frac{1}{2}m \kappa^{\alpha \beta } \eta_{\mu \nu } 
\frac{\partial{X^\mu}}{\partial{x^\alpha}}  
\frac{\partial{X^\nu}}{\partial{x^\beta}}
\label{L}
\end{equation}
where $m$ is the mass of the quantum observer O1, $x^{\alpha }$ and $x^{\beta }$ are the coordinates ($t, x, y, z,..$) or e.g. $(t, r, \theta, \phi)$ internal to the observer and ${X^{\mu }}$ are fields representing the observer's centre of mass position in the reference system of another observer O2.  The factor $\kappa$ is given by:
\begin{equation}
\kappa ^{\alpha \beta} = \frac{dx^\alpha}{dT} \frac{dx^\beta}{dT}
\end{equation}

It is possible that the ${X^\mu}$ might instead represent the coordinates of an observed system, rather than of O1, in O2's reference system~\cite{Dance0901}. The ${x^\alpha}$ might represent the observed system's coordinates in O1's reference system.  

Ordinarily, the coordinate transformation between the O1 and O2 reference systems is simply incorporated into a metric, and is not considered to be of further interest. With one or more quantum mechanical observers in the picture, however, the situation changes. Now one or both of the sets ${X^\mu}$ and ${x^\alpha}$ are subject to quantum mechanical minimum uncertainties in observable distance (and time) intervals. So the transformation is now of inherent quantum mechanical interest.  This has been discussed in more detail in e.g.~\cite{Dance0610}.

The simplest string action is the area of the world-sheet of a string propagating through flat space-time, multiplied by a constant tension:-
\begin{equation}
S_{string} = -\frac{T}{2} \int d\tau d\sigma
\eta^{\alpha \beta } \eta_{\mu \nu } 
\frac{\partial{X^\mu}}{\partial{x^\alpha}}
\frac{\partial{X^\nu}}{\partial{x^\beta}},
\label{str}
\end{equation}
where $X^{\mu }$ are the space-time coordinates of a point on the string, and
$x^{\alpha }$, $x^{\beta}$ are members of $(\tau, \sigma)$.

Focusing on the derivatives in Equations~(\ref{L}) and~(\ref{str}), string theory uses the worldsheet parameters ($\tau, \sigma_i$ ) instead of internal O1 coordinates.  \cite{Dance0601} postulated that a physical correspondence can be made between $(t, r, \theta, \phi)$ and $(\tau, \sigma_i)$, as follows. 

It was postulated in~\cite{Dance0601} that an observer will not be able to extract information about parameters with respect to which the first, quantum mechanical observer's initial and final states are symmetric in its own internal coordinate system. At the quantum level, a common symmetry is rotational symmetry. An example is an H atom whose initial and final states both possess a spherically symmetric $1s$ electronic state. To this H atom observer, the angular coordinates $\theta $ and $\phi $ may be meaningless.  If the H atom begins and ends an observation instead with its electron in (say) an excited state, such as a $2p_z$ state, then the H atom has symmetry about an internal $z$ axis; it was postulated that the concept of the polar coordinate $\phi $ may be meaningless to this observer, and that this coordinate does not appear in the Lagrangian density.  The more symmetries an observer possesses, the less the observer may be able to detect. It was also postulated that a simple quantum mechanical observer such as an atom cannot measure the radial coordinate $r$.

In this way, a correspondence was made between the internal observer coordinates $(t, r, \theta, \phi)$ and the string-theory parameters $(\tau, \sigma_i)$, where $\tau = t$, and the $\sigma_i$ of string theory correspond to the angular coordinates that the quantum observer can detect.  It was suggested that there are 1D-observers (that can detect one angular coordinate) and 2D-observers (which can detect two angular coordinates).  The 1D-observers would correspond to strings, and 2D-observers to 2-branes. 
 
The paper~\cite{Dance0901} included the dynamics of a fermion field (perhaps internally in O1), in addition to the centre of mass term. It was suggested there that quantum uncertainties in the transformation between the reference systems of O1 and O2 might require the use of $d$ spinor fields for the fermion term, where $d$ is the number of spacetime dimensions.  With this fermion term, $\mathcal{L}$ becomes:
\begin{equation}
\mathcal{L}  =  \frac{1}{2}m \kappa^{\alpha \beta } \eta_{\mu \nu } 
                \frac{\partial{X^\mu}}{\partial{x^\alpha}}  
                \frac{\partial{X^\nu}}{\partial{x^\beta}} 
                + i\bar{\psi}^\mu \gamma ^{\alpha} \partial_{\alpha} \psi_\mu
\label{EqL} 
\end{equation}
where, as before, the $x^{\alpha}$, $x^{\beta}$ are coordinates in O1's reference system.  The Lagrangian density then has the form of the Polyakov Lagrangian density in superstring theory.

The paper~\cite{Dance1011} extended the discussion to a quantum mechanical observer O1 that is a molecule. The same suggestions applied, as many molecular bonds have rotational symmetries that could provide the same effective dimensional reduction as in~\cite{Dance0601}. It was also suggested that the ${X^\mu}$ can represent the combined coordinates of a number of entities within O1.  In this way, molecular observers might give rise to string theory Lagrangian densities with apparent spacetime dimensionalities greater than 4, for example 10 or 26. In this physical picture, there is no need for compactification, because the coordinates are not those of an underlying spacetime on which dynamics unfold, but are instead the coordinates of a number of entities within the observer O1, moving in a normal spacetime.  It was also suggested that the ${x^\alpha}$ could be extended to a larger set, corresponding to internal coordinates visible to the molecule O1. Such an enlarged ${x^\alpha}$ set could give rise to theories containing $p$-branes with $p>2$.  Again, the ${X^\mu}$ and ${x^\alpha}$ might instead represent the coordinates of an observed system (rather than of O1), in O2's and O1's reference systems respectively. 

This has summarised the basic physical picture described in previous work including~\cite{Dance0601}. The sections below will now extend the discussion to the AdS/CFT correspondence.

\section{The AdS/CFT correspondence - an overview}

This section provides a brief overview of the AdS/CFT correspondence.

The AdS/CFT correspondence is a conjecture that string theory on AdS space times a closed surface corresponds to a conformal field theory on a boundary of the AdS space. The first and best known example of the AdS/CFT correspondence involves Type IIB string theory on Ad$\rm{S}_{\rm{5}} x \rm{S}^{\rm{5}}$, where $\rm{S}^{\rm{5}}$ is a sphere in a 6-dimensional space~\cite{Malda97}.  
 
Ad$\rm{S}_{\rm{5}}$ is an extension of Minkowski space that includes one extra time coordinate and one extra spatial coordinate. A point in Ad$\rm{S}_{\rm{5}}$ has coordinates $(T_1, X_1, X_2, X_3, T_2, X_4)$ and satisfies the constraint:
\begin{equation}
\Sigma_i T_i^2 - \Sigma_i X_i^2 = K^2
\end{equation}
for a positive constant $K^2$.

\section{Observers and the AdS/CFT correspondence}

In this section, we discuss how the present physical picture might give rise to an AdS/CFT correspondence, focusing on the specific case of Ad$\rm{S}_{\rm{5}} x \rm{S}^{\rm{5}}$. We start with $\mathcal{L}$ as set out in Equation~(\ref{L}) above, and we consider a set of three observation stages: the observers O1 and O2 discussed above, and now a third observer O.  For the case of 
Ad$\rm{S}_{\rm{5}} x \rm{S}^{\rm{5}}$, we consider O1, O2 and O to have specific types of observational capabilities and coordinate systems, designed to fit this example.  We postulate a flat background spacetime.

For the argument in this section, the ${X^\mu}$ will represent O1's coordinates in O2's reference system.  We will take the observers O1, O2, and O to have the following coordinate systems:
\begin{itemize} 
\item{O1 has visible to it three internal angular coordinates and two time coordinates.  In the picture above, this corresponds to O1 being a 1D-observer plus a 2D-observer. The 1D-observer may correspond to a sigma bond, and the 2D-observer might correspond to a pi bond in a double bond in (e.g.) a retinal molecule.  Such an O1 might give rise to a theory containing a string theory 3-brane.}
\item{O2 observes two copies of O1, which are at different locations. For each O1, O2 can observe three angular coordinates, two time coordinates, and a radial distance coordinate which is the distance to that O1. O2 might correspond to an early or intermediate stage of sensory processing.} 
\item{O can see two angular coordinates, one time coordinate, and one radial distance coordinate, corresponding to ordinary Minkowski spacetime. O might stand almost in the position of the normal human observer, not perhaps its final observation stage, but perhaps an intermediate or late stage of sensory processing, after the O2 observation stage.}
\end{itemize}
 
To describe how this physical picture might be linked with the AdS/CFT correspondence, we will consider two scenarios. In the first, we describe how O2 perceives the world. In the second, we consider how O describes its observations of the world.
 
\subsection{O2's description of the world}
 
O2 observes two copies of O1, each described by the Lagrangian density $\mathcal{L}$ in equation~(\ref{L}) above.  For each O1, O2 can observe six coordinates.  We postulate the following constraints on the coordinates observed by O2:
\begin{itemize}
\item{One O1 must be at a timelike interval from O2, otherwise no information can be communicated from O1 to O2. We postulate that this corresponds to a constraint:
\begin{equation}
\Sigma_i T_i^2 - \Sigma_i X_i^2 = K^2 
\end{equation}
for a positive constant $K^2$, for one of the O1 observers. This interval is the sum of the intervals for each observing part of the O1; one is a 1D-observer, the other is a 2D-observer, and adding the intervals gives the six terms above. This constraint contributes a factor Ad$\rm{S}_{\rm{5}}$ to the space of coordinates on which O2 makes its observations, provided the timelike interval is a fixed constant. This might make physical sense, because for O2 to make sense of the world, it is simplest for O2 if the time delay in communication beween O1 and O2 is fixed. If there were an unpredictably varying time delay, O2 might find it difficult to process its input in a way that could give useful output.  Note also that O2 adds the intervals from both observing parts of the observer O1. This may be because O2 cannot tell which time coordinate corresponds to which part of the O1, because the relevant electrons are indistinguishable.}

\item{We postulate the existence of a spatial constraint; expressed in Cartesian coordinates in O2's reference system, the constraint is:
\begin{equation}
(X^2 + Y^2 + Z^2)_{\rm{first O1}} + (X^2 + Y^2 + Z^2)_{\rm{second O1}}  =  K^2,
\end{equation}
where $K^2$ is a positive constant. This is $\rm{S}^{\rm{5}}$. This constraint means that the geometric mean, of the distances from O2 to the two O1 observers, must be a fixed constant.  One could propose an alternative constraint, which could correspond with a different closed manifold.  In the constraint equation above, one of the observed spatial coordinates in O2's reference system has been ignored. This corresponds to O2 selecting only those physical coordinates that assist O2's (or a later-stage observer's) purpose to interpret the external world. The sum above also includes coordinates for only one part of each O1, the 1D-observer or the 2D-observer; this may be because this will be enough for O2 to estimate its distance from each O1.}
\end{itemize}
 
The above constraints might enable O2 to extract useful information from the input which it receives. In this picture, O2 sees a string theory governed by $\mathcal{L}$ on Ad$\rm{S}_{\rm{5}} x \rm{S}^{\rm{5}}$.  

\subsection{O's description of the world}
 
We now consider how the observer O describes the world.  O is taken to be an intelligent observer, which might correspond with a later stage of sensory processing.  
 
O observes four coordinates, which are those of the standard Minkowski spacetime. These four coordinates are those which are of use to O.  O does not observe any other coordinates which to it would be superfluous for its purpose. We postulate that O observes a distance coordinate (which may or may not be the same as that seen by O2), one time and two angular coordinates.

Without knowing exactly what the Lagrangian density $\mathcal{L'}$ for its observed world appears to O to be, we postulate that somewhat like the human observer, O's theory of the world will have invariance under global spatial translations, global rotations, and also (unlike the human observer) global dilatations.  These postulates might amount to saying that in determining a theory of the external world, O identifies objects and their motion in a way that is independent of absolute position, absolute orientation and global scaling of apparent sizes. These statements are true for the human observer, except for dilatations; our perceived world does not generally have scale invariance. We assume that O is not the human observer, that is to say O is not the final human observation stage, but rather an intermediate stage in processing input. An apparent global dilation in size of all objects will be due to the objects being closer to the observer O, and O will tend to eliminate this effect in formulating its dynamical theory of the external world. We postulate that O's representation of the world will also be Lorentz invariant.

This is almost the definition of a (globally) conformally invariant theory.  There is one more criterion: that O's picture of the world be invariant under special conformal transformations. These have the form
\begin{equation}
\mathbf{r} \longrightarrow \mathbf{r^\prime} = \frac{ \mathbf{r} + \mathbf{a}r^2 }{ 1 + 2\mathbf{a}.\mathbf{r} + a^2 r^2}
\end{equation}
This can be rewritten in the form
\begin{equation}
\frac{ \mathbf{r'} }{ r'^{2} }  = \frac{ \mathbf{r} }{ r^2 } + \mathbf{a}
\label{special}
\end{equation}
This is the combination of an inversion $\mathbf{r} \longrightarrow \mathbf{r'} = \mathbf{r}/r^2$ followed by a translation $\mathbf{a}$ and again an inversion, as noted in~\cite{Henkel}.
   
In one spatial dimension, it is easy to show that the special conformal transformation corresponds to a lens equation. In one spatial dimension, let us write:
\begin{equation}
\mathbf{r} = (-d, 0, 0)
\end{equation}
\begin{equation}
\mathbf{r'} = (d', 0, 0)
\end{equation}
and Equation~(\ref{special}) becomes
\begin{equation}
\frac{1}{d} + \frac{1}{d'} = \frac{1}{f}
\end{equation}
where $f = 1/a$.  

This is a lens equation.

The special conformal transformations then may be transformations under which an image formed by a lens is mapped to the object in the external world (or vice versa), with the focal length of the lens being variable. But this is exactly what occurs in the human eye, which observes images on its retina with a variable focal length lens. We postulate that O is such an observer, and that its theory of the external world is independent of whether it is dealing directly with external objects or with their images. 

Then O's theory of the world is a (globally) conformally invariant theory in Minkowski spacetime.

So far we have discussed global conformal invariance. It may be that this can be extended to local invariance. For example, the human observer sees a theory of the external world that has local Lorentz invariance and invariance under local translations and rotations, in the form of general relativity. It may be that O - an intermediate or late sensory processing stage - has such a model of the external world. It is also possible that O's theory of the world (embodied in $\mathcal{L'}$) has invariance under local dilatations and local special conformal transformations. Invariance under local dilatations might correspond to an O whose theory of the world does not depend on how close an individual object might be to O, within its overall field of vision.  Invariance under local special conformal transformations might correspond to an O whose theory of the world cancels out the effect of a focal length that varies for different spacetime points in the external world.  This might mean taking account of the effects  of a lens that is not a perfect lens with a single focal length that applies to all its incoming input, but instead a lens with aberration.
 
Then O's theory of the world might also have local conformal invariance in Minkowski spacetime.
 
We note that if some angles are invisible to any observer, they also appear to drop out of the dynamics. The apparent dimensionality of the set of coordinates ${X^\mu}$ will then be reduced by the number of angular coordinates that are invisible. The resulting theory will appear to O to be conformally invariant in the remaining coordinates.

\subsection{AdS/CFT correspondence: comparing perspectives of O2 and O}

We will now compare the perspectives of O2 and O, using the results from the sections above. 

From above, O2 observes a string theory in the space Ad$\rm{S}_{\rm{5}} x \rm{S}^{\rm{5}}$.  The string theory is governed by the Lagrangian density $\mathcal{L}$ given in Equation~(\ref{L}).  We have also seen above that there might be an observer O that observes the same world as perceived by O2, as a conformally invariant theory in ordinary Minkowski spacetime.  The arguments for the existence of such an observer O seem reasonable and in some correspondence with physical reality, but are not a definitive proof.
 
Therefore, we see that the dynamics perceived by the two different observers O2 and O might give rise to a correspondence that somewhat resembles an AdS/CFT correspondence, specifically the Ad$\rm{S}_{\rm{5}} x \rm{S}^{\rm{5}}$ correspondence. We note that the string theory in the AdS/CFT correspondence of~\cite{Malda97} is Type IIB superstring theory. This theory exists on a spacetime with 10 apparent dimensions, which corresponds to the (6+6) size of the two {$X^\mu $} sets after the two constraints are subtracted from the set of independent coordinates. 
    
More generally, there may be a number of correspondences in string theory, such as the AdS/CFT correspondence, which might be alternatively describable in terms of the ways in which different observers perceive the same systems (or other observers) which they observe.  The observers may each be parts in a chain of observation stages, or otherwise.  To obtain such correspondences, various aspects of the arguments above could be varied.  For example, the observed coordinates of O2 and O could be changed, or dynamics of these observers could be included.  

It might be an interesting future project to attempt to match the different types of string theories with corresponding interpretations in the present physical picture, with defined coordinates observed by different types of observers. Each type of string theory might correspond to a specific set of observer capabilities.

\section{Further discussion} 

The paper~\cite{Dance0601} added a simple quantum mechanical observer term to a Lagrangian density, and suggested that the radial dimension $r$ is not observable by such an observer.  It was suggested that such theories may shed light on string theories, e.g. if the string theory parameters $\tau $ and $\sigma_i$ represent the coordinates which the observer can observe, after observer symmetries effectively eliminate some coordinates, internal to the observer, from observation.  

The present paper has extended the discussion to show a possible link between quantum mechanical observers and an example of the AdS/CFT correspondence.  In the present physical picture, the correspondence arises from the ways in which the two different observers O2 and O describe their world (which may be the external world, other stages of observation).
 
Balasubramanian and McGreevy~\cite{Bala0804}, and separately D. Son~\cite{Son0804}, have suggested applying the AdS/CFT correspondence to cold atoms as a mathematical method to assist workers to understand the theory of cold atoms in certain limits, in particular for energies matching a Feshbach resonance. By contrast, the present approach is quite different. It is not restricted to particular atomic or molecular conditions. Our $\mathcal{L}$ is simple and requires consideration only of the transformation between reference systems, at least one of which is a quantum mechanical reference system. (If there is to be any express consideration of the dynamics of the observers, the focus will be on their centre-of-mass dynamics; the central feature of the present picture is the quantum mechanical transformation between the reference systems.)  For our purposes, we have taken the structures of O1 and of other observers to be given and fixed.  This condition could potentially be relaxed in future.  It may also be appropriate to incorporate decoherence mechanisms.

We have above treated $X$ as a vector field over $x$. It may be appropriate more generally in quantum field theory to replace fields over spacetime points $x$ by fields over quantum fields $X(x)$. For example, $\phi(x)$ could be replaced by $\phi(X(x))$ in Lagrangian densities. This might be an approximation, if both the $X$ and $x$ coordinates (in the reference systems of two observers) are subject to quantum uncertainty.

If O2 and O represent different stages in processing of sensory input, for example in a visual system, it may be worthwhile to note that a stage such as O2, observing two separate observers O1, may be necessary for later observation stages to be able to observe a depth or distance coordinate.  In the light of this, it appears reasonable for O2 to have a description such as that outlined above.  Reversing this logic, it may be that correspondences such as the AdS/CFT correspondence might even be able to shed light on the operation of sensory or other observation systems.
 
\section{Conclusion}

The present paper has outlined how consideration of the reference systems of different observers, at least one of which is quantum mechanical, might provide insights into the nature of the AdS/CFT correspondence.  We have here suggested that an AdS/CFT correspondence may arise from the ways in which two different observers describe the same observed systems. The observed systems may be in the external world or may be other observers (such as O1) earlier in the chain of observation. We have suggested that other such correspondences might arise as different forms of theories about the external world, or other observed systems, as perceived by different observers.  There are a number of potentially interesting avenues for future work in this area.

\end{document}